\documentclass[twocolumn,showpacs,amsmath,amssymb,superscriptaddress]{revtex4}
\usepackage{graphicx}
\usepackage{dcolumn}
\usepackage{bm}

\begin{document}

\title{Surface properties of neutron-rich exotic
nuclei: A source for studying the nuclear symmetry energy}

\author{M.K. Gaidarov}
\affiliation{Institute for Nuclear Research and Nuclear Energy,
Bulgarian Academy of Sciences, Sofia 1784, Bulgaria}

\author{A.N. Antonov}
\affiliation{Institute for Nuclear Research and Nuclear Energy,
Bulgarian Academy of Sciences, Sofia 1784, Bulgaria}

\author{P. Sarriguren}
\affiliation{Instituto de Estructura de la Materia, IEM-CSIC, Serrano 123, E-28006 Madrid, Spain}

\author{E. Moya de Guerra}
\affiliation{Departamento de Fisica Atomica, Molecular y Nuclear,
Facultad de Ciencias Fisicas, Universidad Complutense de Madrid,
E-28040 Madrid, Spain}


\begin{abstract}
We study the correlation between the thickness of the neutron skin
in finite nuclei and the nuclear symmetry energy for isotopic
chains of even-even Ni, Sn, and Pb nuclei in the framework of the
deformed self-consistent mean-field Skyrme HF+BCS method. The
symmetry energy, the neutron pressure and the asymmetric
compressibility in finite nuclei are calculated within the
coherent density fluctuation model using the symmetry energy as a
function of density within the Brueckner energy-density
functional. The mass dependence of the nuclear symmetry energy and
the neutron skin thickness are also studied together with the role
of the neutron-proton asymmetry. A correlation between the
parameters of the equation of state (symmetry energy and its
density slope) and the neutron skin is suggested in the isotopic
chains of Ni, Sn, and Pb nuclei.
\end{abstract}


\maketitle

\section{Introduction}

The nuclear symmetry energy is a quantity of crucial importance in
different areas of nuclear physics, including structure of
ground-state nuclei \cite{Niksic2008,Giai2010,Dalen2010}, dynamics
of heavy-ion reactions \cite{Li2008,Chen2008,Colonna2009}, physics
of giant collective excitations \cite{Rodin2007} and physics of
neutron stars \cite{Steiner2005,Psonis2007,Sharma2009}. Recently,
the interest in the symmetry energy has been stirred up by novel
astrophysical observations and by the availability of exotic beams
in accelerators that provide additional information to the
standard nuclear asymmetry studies based on stable nuclei.
Particularly important in the different areas, and similarly
uncertain, is the density dependence of the symmetry energy in
uniform matter. As can be seen e.g., in
Refs.~\cite{Li2006,Piekarewicz2009,Vidana2009,Sammarruca2009,Dan2010},
an increasing wide range of theoretical conclusions are being
proposed on that density dependence, as well as on some associated
nuclear characteristics. In the last years, the temperature
dependence of single-particle properties in nuclear and neutron
matter was also broadly investigated (e.g.,
Refs.~\cite{Moustakidis2007,Sammarruca2008}).

Measurements of nuclear masses, densities, and collective
excitations have allowed to resolve some of the basic features of
the equation of state (EOS) of nuclear matter. However, the
asymmetry properties of the EOS due to different neutron and
proton numbers remain more elusive to date, and the study of the
isospin dependent properties of asymmetric nuclear matter
\cite{Oyamatsu98,Zuo99,Roy2009,Chen2009,Gogelein2009} and the
density dependence of the nuclear symmetry energy remains a prime
objective. The new radioactive ion beam facilities at CSR (China),
FAIR (Germany), RIKEN (Japan), SPIRAL2/GANIL (France) and the
upcoming FRIB (USA) will provide the possibility of exploring the
properties of nuclear matter and nuclei under the extreme
condition of large isospin asymmetry.

Experimentally, the symmetry energy is not a directly measurable
quantity and has to be extracted indirectly from observables that
are related to it (see, for example, the recent review
\cite{Shetty2010}). The neutron skin thickness of nuclei is a
sensitive probe of the nuclear symmetry energy, although its
precise measurement is difficult to obtain. At present neutron
skin thicknesses are derived from pygmy dipole resonances
measurements \cite{Klimk2007}, data from antiprotonic atoms
\cite{Centelles2009} and other methods for its extraction like
reactions and giant resonances. This allows one to constrain the
parameters describing the nuclear symmetry energy.

Considering nuclei from the isotopic chains of Sn and Pb, Warda
{\it et al.} \cite{Warda2010} studied theoretically the bulk and
the surface nature of the formation of the neutron skin, a concept
that can be applied when analyzing the experimental data. The same
authors indicated in Ref.~\cite{Centelles2010} the role of the
stiffness of the nuclear symmetry energy on the origin of the
neutron skin thickness of $^{208}$Pb, the latter being decomposed
into bulk and surface components. Also, Danielewicz \cite{Dan2003}
has demonstrated that the ratio of the volume symmetry energy to
the surface symmetry energy is closely related to the neutron skin
thickness. The correlation between bulk and surface symmetry
energy was further discussed in
Refs.~\cite{Steiner2005,Danielewicz,Diep2007,Kolomietz,Nikolov2011}.
Recent calculations of thermal nuclear properties showed that the
surface symmetry energy term is more sensitive to temperature than
the volume energy term \cite{Lee2010}.

The neutron skin thickness, generally defined as the difference
between neutron and proton rms radii in the atomic nucleus, is
closely correlated with the density dependence of the nuclear
symmetry energy and with the equation of state of pure neutron
matter. Moreover, it has been shown that the neutron skin
thickness in heavy nuclei, like $^{208}$Pb, calculated in
mean-field models with either nonrelativistic or relativistic
effective nuclear interactions, displays a linear correlation with
the slope of the neutron EOS obtained with the same interactions
at a neutron density $\rho \approx 0.10$ fm$^{-3}$
\cite{Brown2000,Typel2001}. The statistical analysis performed in
Ref.~\cite{Reinhard2010} clearly shows the strong correlation that
exists between the neutron form factor of $^{208}$Pb and isovector
indicators such as neutron skins and radii in neutron-rich nuclei,
dipole polarizability, and nuclear matter properties (symmetry
energy and pressure).

The symmetry energy of finite nuclei at saturation density is
often extracted by fitting ground state masses with various
versions of the liquid-drop mass (LDM) formula within liquid-drop
models \cite{Myers66,Moller95,Pomorski2003}. It has been also
studied in the random phase approximation based on the
Hartree-Fock (HF) approach \cite{Carbone2010} or effective
relativistic Lagrangians with density-dependent meson-nucleon
vertex functions \cite{Vretenar2003}, energy density functionals
of Skyrme force \cite{Chen2005,Yoshida2006,Chen2010} as well as
relativistic nucleon-nucleon interaction \cite{Lee98,Agrawal2010}.
In the present work the symmetry energy will be studied {\it in a
wide range of finite nuclei} on the basis of the Brueckner
energy-density functional for nuclear matter
\cite{Brueckner68,Brueckner69} and using the coherent density
fluctuation model (CDFM) (e.g., Refs.~\cite{Ant80,AHP}). The
latter is a natural extension of the Fermi gas model based on the
generator coordinate method \cite{AHP,Grif57} and includes
long-range correlations of collective type. The CDFM has been
applied to calculate the energies, the density distributions and
rms radii of the ground and first monopole states in $^{4}$He,
$^{16}$O and $^{40}$Ca nuclei \cite{Ant88}. An analysis of
breathing monopole states whose description is related mainly to
general characteristics of atomic nuclei and weakly to the
peculiarities of nuclear structure in the framework of the CDFM
has been performed and also results for the incompressibility of
finite nuclei have been reported \cite{AHP,Ant91}. In
Refs.~\cite{Ant2004,Ant2005} the CDFM has been employed to
calculate the scaling function in nuclei using the relativistic
Fermi gas scaling function and the result has been applied to
lepton scattering processes
\cite{Ant2004,Ant2005,Ant2006a,Ant2006b,Ant2008,Ant2007,Ant2009}.
The CDFM analyses became useful to obtain information about the
role of the nucleon momentum and density distributions for the
explanation of superscaling in lepton-nucleus scattering
\cite{Ant2005,Ant2006a}. The CDFM scaling function has been used
to predict cross sections for several processes: inclusive
electron scattering in the quasielastic and $\Delta$ regions
\cite{Ant2006b,Ant2008} and neutrino (antineutrino) scattering
both for charge-changing \cite{Ant2008,Ant2009} and for
neutral-current \cite{Ant2007,Ant2009} processes.  Recently, the
CDFM was applied to study the scaling function and its connection
with the spectral function and the momentum distribution
\cite{Cab2010}.

In our previous study \cite{Sarriguren2007}, we analyzed within a
deformed HF+BCS approach with Skyrme-type density-dependent
effective interactions the existence of a skin in nuclei far from
the stability line by testing various definitions previously
proposed \cite{Mizutori2000,Fukunishi93}. We have found that the
neutron skin determined by the difference between neutron and
proton radii using diffraction parameters defined in the Helm
model \cite{Mizutori2000} exhibits a more smooth, gradual increase
with the neutron excess than the one obtained by means of
definition from Ref.~\cite{Fukunishi93}. As a result, the absolute
size of the skin ranges from 0.4 fm when the Helm model is used up
to almost 1 fm for the heaviest isotopes using the criteria from
Ref.~\cite{Fukunishi93}. The same Helm model definitions of the
neutron skin thickness were studied in Ref.~\cite{Schunck2008} in
spherical self-consistent Hartree-Fock-Bogoliubov (HFB)
calculations with finite-range Gogny forces. We would like to
emphasize the consistency of the results of both approaches for
the case of Ni and Sn isotopes \cite{Sarriguren2007,Schunck2008}.
Some particular investigations of the neutron skin thickness of
neutron-rich nuclei and its connections with the symmetry energy
have been performed in the droplet model with surface width
dependence \cite{Warda2009} and in studies based on the
Landau-Migdal approximation \cite{Diep2003}.

In the present work we investigate the relation between the
neutron skin thickness and some nuclear matter properties in
finite nuclei, such as the symmetry energy at the saturation
point, symmetry pressure (proportional to the slope of the bulk
symmetry energy), and asymmetric compressibility, considering
nuclei in given isotopic chains and within a certain theoretical
approach. In addition to various linear relations between several
quantities in bulk matter and for a given nucleus that have been
observed and tested within different theoretical methods (e.g.
nonrelativistic calculations with different Skyrme parameter sets
and relativistic models), we are looking forward to establish a
possible correlation between the skin thickness and these
quantities and to clarify to what extent this correlation is
appropriate for a given isotopic chain. As in our previous papers
\cite{Sarriguren2007,Antonov2005}, we choose some medium and heavy
Ni and Sn isotopes, because, first, they are primary candidates to
be measured at the upcoming experimental facilities, and second,
for them several predictions have been made concerning the nuclear
skin emergence. In addition, we present some results for a chain
of Pb isotopes being inspired by the significant interest (in both
experiment \cite{prex} and theory \cite{Moreno2010}) to study, in
particular, the neutron distribution of $^{208}$Pb and its rms
radius. The densities of these nuclei were calculated within a
deformed HF+BCS approach with Skyrme-type density-dependent
effective interactions \cite{vautherin,Guerra91}. The results
obtained in Ref.~\cite{Sarriguren2007} demonstrated the ability of
our microscopic theoretical method to predict the nuclear skin in
exotic nuclei. As already mentioned, for a link between nuclear
matter and finite nuclei we adopt the energy-density functional of
Brueckner {\it et al.} \cite{Brueckner68,Brueckner69} that has
been able to reproduce the binding energies and rms radii of
light, medium, and heavy nuclei. The intrinsic properties of these
nuclei are finally calculated in the CDFM framework that provides
{\it an extension to realistic finite nuclear systems}. The
evolution of the symmetry energy as we increase the number of
neutrons is also studied. Additionally to
Ref.~\cite{Sarriguren2007}, we pay more attention to the role of
the neutron-proton asymmetry on the neutron skin formation.

The structure of this article is the following. In Sec.~II we
present the common definitions of symmetry energy and properties
of nuclear matter which characterize its density dependence around
normal nuclear matter density. Section~III contains the CDFM
formalism that provides a way to calculate the intrinsic
quantities in finite nuclei. There we give also the basic
expressions for the model density distributions and nuclear radii
obtained in the deformed HF+BCS method. The numerical results and
discussions are presented in Sec.~IV. We draw the main conclusions
of this study in Sec. V.

\section{The key EOS parameters in nuclear matter}

The symmetry energy $S(\rho)$ is defined by a Taylor series
expansion of the energy per particle for nuclear matter (NM) in
terms of the isospin asymmetry $\delta=(\rho_{n}-\rho_{p})/\rho$
\begin{equation}
E(\rho,\delta)=E(\rho,0)+S(\rho)\delta^2+O(\delta^4)+
\cdot\cdot\cdot \;\; ,
\label{eq:1}
\end{equation}
where $\rho=\rho_{n}+\rho_{p}$ is the baryon density with
$\rho_{n}$ and $\rho_{p}$ denoting the neutron and proton
densities, respectively (see, e.g. \cite{Diep2003,Chen2011}). Odd
powers of $\delta$ are forbidden by the isospin symmetry and the
terms proportional to $\delta^{4}$ and higher orders are found to
be negligible.

Near the saturation density $\rho_{0}$ the energy of isospin-symmetric matter,
$E(\rho,0)$, and the symmetry energy, $S(\rho)$, can be expanded as
\begin{equation}
E(\rho,0)=E_{0}+\frac{K}{18\rho_{0}^{2}}(\rho-\rho_{0})^{2}+
\cdot\cdot\cdot \;\;  ,
\label{eq:2}
\end{equation}
and
\begin{eqnarray}
S(\rho)&=&\frac{1}{2}\left.
\frac{\partial^{2}E(\rho,\delta)}{\partial\delta^{2}} \right
|_{\delta=0} \nonumber \\ 
&=& \left. a_{4}+\frac{p_{0}}{\rho_{0}^{2}}(\rho-\rho_0)+\frac{\Delta
K}{18\rho_{0}^{2}}(\rho-\rho_{0})^{2}+ \right. \cdot\cdot\cdot \;.
\label{eq:3}
\end{eqnarray}
The parameter $a_{4}$ is the symmetry energy at equilibrium
($\rho=\rho_{0}$). The pressure $p_{0}^{NM}$
\begin{equation}
p_{0}^{NM}=\rho_{0}^{2}\left. \frac{\partial{S}}{\partial{\rho}}
\right |_{\rho=\rho_{0}}
\label{eq:4}
\end{equation}
and the curvature $\Delta K^{NM}$
\begin{equation}
\Delta K^{NM}=9\rho_{0}^{2}\left.
\frac{\partial^{2}S}{\partial\rho^{2}} \right |_{\rho=\rho_{0}}
\label{eq:5}
\end{equation}
of the nuclear symmetry energy at $\rho_{0}$ govern its density
dependence and thus provide important information on the
properties of the nuclear symmetry energy at both high and low
densities. The widely used "slope" parameter $L^{NM}$ is related
to the pressure $p_{0}^{NM}$ [Eq.~(\ref{eq:4})] by
\begin{equation}
L^{NM}=\frac{3p_{0}^{NM}}{\rho_{0}}.
\label{eq:6}
\end{equation}

We remark that our present knowledge of these basic properties of
the symmetry term around saturation is still very poor (see the
analysis in Ref.~\cite{Furnstahl2002} and references therein). In
particular we note the uncertainty of the symmetry pressure at
$\rho_{0}$ (sometimes it can vary by a factor of three) which is
important for structure calculations. In practice, predictions for
the symmetry energy vary substantially: {\it e.g.}, $a_{4}\equiv
S(\rho_{0})=28-38$ MeV. An empirical value of $a_{4}\approx 29$
MeV has been extracted with reasonable accuracy from finite nuclei
by fitting ground-state energies using the generalized
Weizs\"{a}cker mass formula \cite{Danielewicz}:
\begin{eqnarray}
E(N,Z)&=&E_{mac}+E_{mic}\nonumber \\
&=&E_{V}+E_{S}+E_{a}+E_{C}+E_{mic} \nonumber \\
&=&-a_{V}A+a_{S}A^{2/3}+a_{a}\frac{(N-Z)^{2}}{A} \nonumber \\
&+&a_{C}\frac{Z^{2}}{A^{1/3}}+E_{mic},
\label{eq:mass}
\end{eqnarray}
where $N$ and $Z$ are the neutron and proton numbers, respectively, and $A=N+Z$
is the mass number. In Eq.~(\ref{eq:mass}) $a_{V}$, $a_{S}$, $a_{a}$ and $a_{C}$ are
the volume, surface, symmetry and Coulomb coefficients. The microscopic energy $E_{mic}$
contains shell- and pairing-energy corrections. It is worth to mention that
the volume component of the symmetry energy coefficient $a_{a}$ (or volume-symmetry
coefficient in the empirical mass formula (\ref{eq:mass})) has to be attributed
to the parameter $a_{4}$ in Eq.~(\ref{eq:3}).

\section{Symmetry energy parameters of finite nuclei in CDFM}

The CDFM was suggested and developed in Refs.~\cite{Ant80,AHP}.
The model is related to the delta-function limit of the generator
coordinate method \cite{Grif57,AHP}. It is shown in the model that
the one-body density matrix of the nucleus $\rho({\bf r},{\bf
r^{\prime}})$ can be written as a coherent superposition of the
one-body density matrices (OBDM) for spherical "pieces" of nuclear
matter (so-called "fluctons") with densities
\begin{equation}
\rho_{x}({\bf r})=\rho_{0}(x)\Theta(x-|{\bf r}|),
\label{eq:7}
\end{equation}
where
\begin{equation}
\rho_{0}(x)=\frac{3A}{4\pi x^{3}}.
\label{eq:8}
\end{equation}
The generator coordinate $x$ is the radius of a sphere containing
Fermi gas of all $A$ nucleons uniformly distributed in it. It is
appropriate to use for such a system OBDM of the form:
\begin{eqnarray}
\rho_{x}({\bf r},{\bf r^{\prime}})&=&3\rho_{0}(x)
\frac{j_{1}(k_{F}(x)|{\bf r}-{\bf r^{\prime}}|)}{(k_{F}(x)|{\bf
r}-{\bf r^{\prime}}|)}\nonumber \\  & \times & \Theta \left
(x-\frac{|{\bf r}+{\bf r^{\prime}}|}{2}\right ),
\label{eq:9}
\end{eqnarray}
where $j_{1}$ is the first-order spherical Bessel function and
\begin{equation}
k_{F}(x)=\left(\frac{3\pi^{2}}{2}\rho_{0}(x)\right )^{1/3}\equiv
\frac{\alpha}{x}
\label{eq:10}
\end{equation}
with
%
\begin{equation}
\alpha=\left(\frac{9\pi A}{8}\right )^{1/3}\simeq 1.52A^{1/3}
\label{eq:11}
\end{equation}
is the Fermi momentum of such a formation. Then the OBDM for the
finite nuclear system can be written as a superposition of the
OBDM's from Eq.~(\ref{eq:9}):
\begin{equation}
\rho({\bf r},{\bf r^{\prime}})=\int_{0}^{\infty}dx |{\cal
F}(x)|^{2} \rho_{x}({\bf r},{\bf r^{\prime}}).
\label{eq:12}
\end{equation}
In the CDFM the Wigner distribution function which corresponds to
the OBDM from Eq. (\ref{eq:12}) is:
\begin{equation}
W({\bf r},{\bf k})=\int_{0}^{\infty}dx|{\cal F}(x)|^{2} W_{x}({\bf
r},{\bf k}),
\label{eq:13}
\end{equation}
where
\begin{equation}
W_{x}({\bf r},{\bf k})=\frac{4}{(2\pi)^{3}}\Theta (x-|{\bf
r}|)\Theta (k_{F}(x)-|{\bf k}|).
\label{eq:14}
\end{equation}
Correspondingly to $W({\bf r},{\bf k})$ from Eq.~(\ref{eq:13}),
the density $\rho({\bf r})$ in the CDFM is expressed by means of
the same weight function $|{\cal F}(x)|^{2}$:
\begin{eqnarray}
\rho({\bf r})&=&\int d{\bf k}W({\bf r},{\bf k}) \nonumber \\
&=&\int_{0}^{\infty}dx|{\cal F}(x)|^{2}\frac{3A}{4\pi x^{3}}\Theta
(x-|{\bf r}|)
\label{eq:15}
\end{eqnarray}
normalized to the mass number:
\begin{equation}
\int \rho({\bf r})d{\bf r}=A.
\label{eq:16}
\end{equation}
If one takes the delta-function approximation to the Hill-Wheeler
integral equation in the generator coordinate method one gets a
differential equation for the weight function ${\cal F}(x)$
\cite{AHP,Grif57}. Instead of solving this differential equation
we adopt a convenient approach to the weight function $|{\cal
F}(x)|^{2}$ proposed in Refs.~\cite{Ant80,AHP}. In the case of
monotonically decreasing local densities ({\it i.e.} for
$d\rho(r)/dr\leq 0$), the latter can be obtained by means of a
known density distribution $\rho(r)$ for a given nucleus [from Eq.
(\ref{eq:15})]:
\begin{equation}
|{\cal F}(x)|^{2}=-\frac{1}{\rho_{0}(x)} \left.
\frac{d\rho(r)}{dr}\right |_{r=x} .
\label{eq:17}
\end{equation}
The normalization of the weight function is:
\begin{equation}
\int_{0}^{\infty}dx |{\cal F}(x)|^{2}=1.
\label{eq:18}
\end{equation}
%
%
%
%
%
%

Considering the pieces of nuclear matter with density
$\rho_{0}(x)$ (\ref{eq:8}) one can use for the matrix element
$V(x)$ of the nuclear Hamiltonian the corresponding nuclear matter
energy from the method of Brueckner {\it et al.}
\cite{Brueckner68,Brueckner69}. In this energy-density method the
expression for $V(x)$ reads
\begin{equation}
V(x)=A V_{0}(x)+V_{C}-V_{CO},
\label{eq:19}
\end{equation}
where
\begin{eqnarray}
V_{0}(x)&=&37.53[(1+\delta)^{5/3}+(1-\delta)^{5/3}]\rho_{0}^{2/3}(x)\nonumber \\
&+&b_{1}\rho_{0}(x)+b_{2}\rho_{0}^{4/3}(x)+b_{3}\rho_{0}^{5/3}(x)\nonumber \\
&+&\delta^{2}[b_{4}\rho_{0}(x)+b_{5}\rho_{0}^{4/3}(x)+b_{6}\rho_{0}^{5/3}(x)]
\label{eq:20}
\end{eqnarray}
with
\begin{eqnarray}
b_{1}&=&-741.28, \;\;\; b_{2}=1179.89, \;\;\; b_{3}=-467.54,\nonumber \\
b_{4}&=&148.26, \;\;\;\;\;\; b_{5}=372.84, \;\;\;\; b_{6}=-769.57.
\label{eq:21}
\end{eqnarray}

$V_{0}(x)$ in Eq.~(\ref{eq:19}) corresponds to the energy per
nucleon in nuclear matter (in MeV) with the account for the
neutron-proton asymmetry. $V_{C}$ is the Coulomb energy of protons
in a "flucton":
\begin{equation}
V_{C}=\frac{3}{5} \frac{Z^{2}e^{2}}{x}
\label{eq:22}
\end{equation}
and the Coulomb exchange energy is:
\begin{equation}
V_{CO}=0.7386 Ze^{2} (3Z/4\pi x^{3})^{1/3}.
\label{eq:23}
\end{equation}

Thus, in the Brueckner EOS [Eq.~(\ref{eq:20})], the potential
symmetry energy turns out to be proportional to $\delta^{2}$. Only
in the kinetic energy the dependence on $\delta$ is more
complicated. Substituting $V_{0}(x)$ in Eq.~(\ref{eq:3}) and
taking the second derivative, the symmetry energy $S^{NM}(x)$ of
the nuclear matter with density $\rho_{0}(x)$ (the coefficient
$a_{4}$ in Eq.~(\ref{eq:3})) can be obtained:
\begin{eqnarray}
S^{NM}(x)&=&41.7\rho_{0}^{2/3}(x)+b_{4}\rho_{0}(x) \nonumber \\
&+&b_{5}\rho_{0}^{4/3}(x)+b_{6}\rho_{0}^{5/3}(x).
\label{eq:24}
\end{eqnarray}
The corresponding analytical expressions for the pressure
$p_{0}^{NM}(x)$ and asymmetric compressibility $\Delta K^{NM}(x)$
of such a system in the Brueckner theory have the form:
\begin{eqnarray}
p_{0}^{NM}(x)&=&27.8\rho_{0}^{5/3}(x)+b_{4}\rho_{0}^{2}(x) \nonumber \\
&+&\frac{4}{3}b_{5}\rho_{0}^{7/3}(x)+\frac{5}{3}b_{6}\rho_{0}^{8/3}(x)
\label{eq:24a}
\end{eqnarray}
and
\begin{eqnarray}
\Delta
K^{NM}(x)&=&-83.4\rho_{0}^{2/3}(x)+4b_{5}\rho_{0}^{4/3}(x)\nonumber
\\
&+&10b_{6}\rho_{0}^{5/3}(x).
\label{eq:24b}
\end{eqnarray}

Our basic assumption within the CDFM is that the symmetry energy,
the slope and the curvature for finite nuclei can be defined
weighting these quantities for nuclear matter (with a given
density $\rho_{0}(x)$ (\ref{eq:8})) by means of the weight
function $|{\cal F}(x)|^{2}$. Thus, in the CDFM they will be an
infinite superposition of the corresponding nuclear matter
quantities. Following the CDFM scheme, the symmetry energy, the
slope and the curvature have the following forms:
\begin{equation}
s=\int_{0}^{\infty}dx|{\cal F}(x)|^{2}S^{NM}(x),
\label{eq:25}
\end{equation}
\begin{equation}
p_{0}=\int_{0}^{\infty}dx|{\cal F}(x)|^{2}p_{0}^{NM}(x),
\label{eq:26}
\end{equation}
\begin{equation}
\Delta K=\int_{0}^{\infty}dx|{\cal F}(x)|^{2}\Delta K^{NM}(x).
\label{eq:27}
\end{equation}
We note that in the limit case when $\rho(r)=\rho_{0}\Theta (R-r)$
($\rho_{0}$ being the saturation density of symmetric nuclear
matter) and $|{\cal F}(x)|^{2}$ becomes a delta-function (see
Eq.~(\ref{eq:17})), Eq.~(\ref{eq:25}) reduces to
$S^{NM}(\rho_{0})=a_{4}$.

The results that will be shown and discussed in the next section
have been obtained from self-consistent deformed Hartree-Fock
calculations with density-dependent Skyrme interactions
\cite{vautherin} and accounting for pairing correlations. Pairing
between like nucleons has been included by solving the BCS
equations at each iteration with a fixed pairing strength that
reproduces the odd-even experimental mass differences.

We consider in this paper the Skyrme SLy4 \cite{sly4}, Sk3
\cite{sk3} and  SG2 \cite{sg2} parametrizations, namely because
they are among the most extensively used Skyrme forces and have
been already used in our previous paper \cite{Sarriguren2007}. In
addition, we probe a recent LNS parameter set from
Ref.~\cite{Cao2006}, derived in the framework of the
Bruekner-Hartree-Fock approximation that has been shown to work
reasonably well for describing finite nuclei in the HF
approximation.

The spin-independent proton and neutron densities are given by
\cite{Sarriguren2007,Guerra91}
\begin{equation}
\rho({\vec R})=\rho (r,z)=\sum _{i} 2v_i^2\rho_i(r,z)\, ,
\label{eq:28}
\end{equation}
in terms of the occupation probabilities $v_i^2$ resulting from
the BCS equations and the single-particle densities $\rho_i$
\begin{equation}
\rho_i({\vec R})=  \rho_i(r,z)=|\Phi^+_i(r,z)|^2+
|\Phi^-_i(r,z)|^2 \, ,
\label{eq:29}
\end{equation}
with
\begin{eqnarray}
\Phi^\pm _i(r,z)&=&{1\over \sqrt{2\pi}}\nonumber \\
&\times & \sum_{\alpha}\, \delta_{\Sigma, \pm 1/2}\,
\delta_{\Lambda,\Lambda^\mp}\, C_\alpha ^i\, \psi_{n_r}^\Lambda
(r) \, \psi_{n_z}(z)
\label{eq:30}
\end{eqnarray}
and $\alpha=\{n_r,n_z,\Lambda,\Sigma\}$. The functions
$\psi^{\Lambda}_{n_r}(r)$ and $\psi_{n_z}(z)$ entering
Eq.~(\ref{eq:30}) are defined in terms of Laguerre and Hermite
polynomials
\begin{equation}
\psi^{\Lambda}_{n_r}(r)=\sqrt{\frac{n_r}{(n_r+\Lambda )!}} \,
\beta_{\perp}\, \sqrt{2}\, \eta^{\Lambda/2}\, e^{-\eta/2}\,
L_{n_r}^{\Lambda}(\eta) \, ,
\end{equation}
\begin{equation}
\psi_{n_z}(z)= \sqrt{\frac{1}{\sqrt{\pi}2^{n_z}n_z!}} \,
\beta^{1/2}_z\, e^{-{\xi}^2/2}\, H_{n_z}(\xi) \, ,
\end{equation}
with
\begin{eqnarray}
\beta_z=(m\omega_z/\hbar )^{1/2}&,&\quad
\beta_\perp=(m\omega_\perp/\hbar )^{1/2},\nonumber \\
\quad \xi=z\beta_z&,&\quad \eta=r^2\beta_\perp ^2 \, .
\end{eqnarray}
The normalization of the densities is given by
\begin{equation}
\int \rho({\vec R}) d{\vec R} = X,
\label{eq:31}
\end{equation}
with $X=Z,\, N$ for protons and neutrons, respectively.

The mean square radii for protons and neutrons are defined as
\begin{equation}
<r_{\rm p,n}^2> =\frac{ \int R^2\rho_{\rm p,n}({\vec R})d{\vec R}}
{\int \rho_{\rm p,n}({\vec R})d{\vec R}} \, ,
\label{eq:32}
\end{equation}
and the rms radii are given by
\begin{equation}
r_{\rm p,n}=<r_{\rm p,n}^2> ^{1/2} \, .
\label{eq:33}
\end{equation}
Then the neutron skin thickness is usually characterized by the
difference of neutron and proton rms radii:
\begin{equation}
\Delta R=<r_{\rm n}^2> ^{1/2}-<r_{\rm p}^2> ^{1/2}.
\label{eq:34}
\end{equation}

\section{Results and discussion}

\begin{figure*}
\centering
\includegraphics[width=170mm]{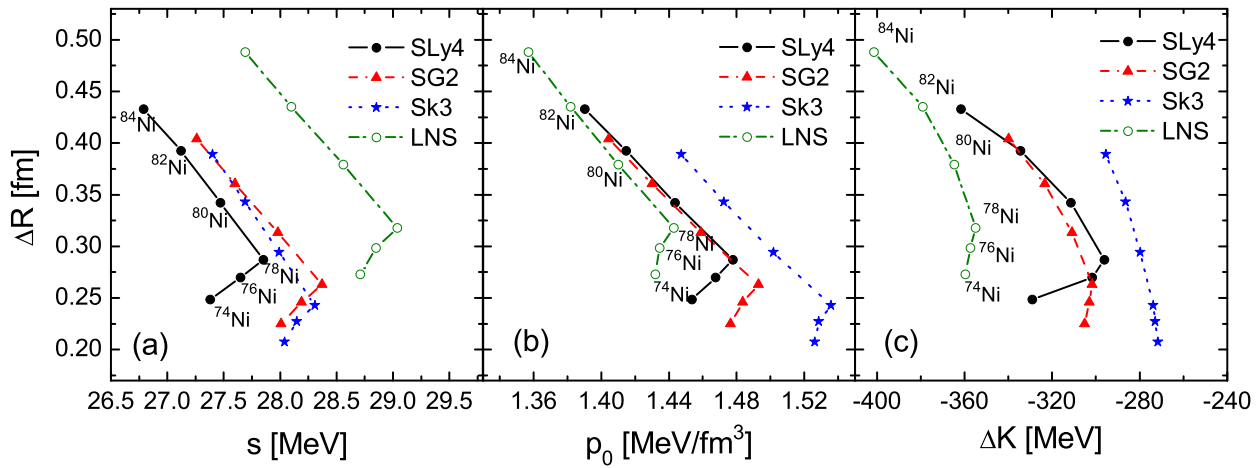}
\caption[]{(Color online) HF+BCS neutron skin thicknesses $\Delta
R$ for Ni isotopes as a function of the symmetry energy $s$ (a),
pressure $p_{0}$ (b), and asymmetric compressibility $\Delta K$
(c) calculated with SLy4, SG2, Sk3, and LNS forces.
\label{fig1}}
\end{figure*}

As the main emphasis of the present study is to inspect the
correlation of the neutron skin thickness $\Delta R$ of nuclei in
a given isotopic chain with the $s$, $p_{0}$ and $\Delta K$
parameters extracted from the density dependence of the symmetry
energy around saturation density, we show first in Fig.~\ref{fig1}
the results for Ni isotopes. The symmetry energy, the pressure and
the asymmetric compressibility are calculated within the CDFM
according to Eqs.~(\ref{eq:25})-(\ref{eq:27}) by using the weight
functions (\ref{eq:17}) calculated from the self-consistent
densities in Eq.~(\ref{eq:28}). The differences between the
neutron and proton rms radii of these isotopes [Eq.~(\ref{eq:34})]
are obtained from HF+BCS calculations using four different Skyrme
forces, SLy4, SG2, Sk3 and LNS. It is seen from Fig.~\ref{fig1}(a)
that there exists an approximate linear correlation between
$\Delta R$ and $s$ for the even-even Ni isotopes with $A=74-84$.
We observe a smooth growth of the symmetry energy till the
double-magic nucleus $^{78}$Ni ($N=50$) and then a linear decrease
of $s$ while the neutron skin thickness of the isotopes increases.
This behavior is valid for all Skyrme parametrizations used in the
calculations, in particular, the average slope of $\Delta R$ for
various forces is almost the same. The LNS force yields larger
values of $s$ comparing to the other three Skyrme interactions. In
this case the small deviation can be attributed to the fact that
the LNS force has not been fitted to finite nuclei and therefore,
one cannot expect a good quantitative description at the same
level as purely phenomenological Skyrme forces \cite{Cao2006}. As
a consequence, the neutron skin thickness calculated with LNS
force has larger size with respect to the other three forces whose
results for $\Delta R$ are comparable with each other.
Nevertheless, it is worth to compare its predictions not only for
ground-state nuclear properties but also for EOS characteristics
of finite systems with those of other commonly used Skyrme forces.

We also find a similar approximate linear correlation for Ni
isotopes between $\Delta R$ and $p_{0}$ [Fig.~\ref{fig1}(b)] and
less strong correlation between $\Delta R$ and $\Delta K$
[Fig.~\ref{fig1}(c)]. As in the symmetry energy case, the behavior
of the curves drawn in these plots shows the same tendency, namely
the inflexion point transition at the double-magic $^{78}$Ni
nucleus. We would like to note that the predictions for the
difference $\Delta R$ between the rms radii of neutrons and
protons with Skyrme forces obtained in Ref.~\cite{Sarriguren2007}
exhibited a steep change at the same place in which the number of
neutrons starts to increase in the chain of nickel isotopes. In
addition, the calculated results for the two-neutron separation
energies and neutron and matter rms radii of the even Ni isotopes
obtained in the relativistic Hartree-Bogoliubov framework
\cite{Meng98} showed a quite strong kink at $N=50$. Also one can
see from Figs.~\ref{fig1}(b) and \ref{fig1}(c) that the calculated
values for $p_{0}$ and $\Delta K$ are lower in the case of LNS
force than for the other three Skyrme parameter sets.

\begin{figure}
\centering
\includegraphics[width=85mm]{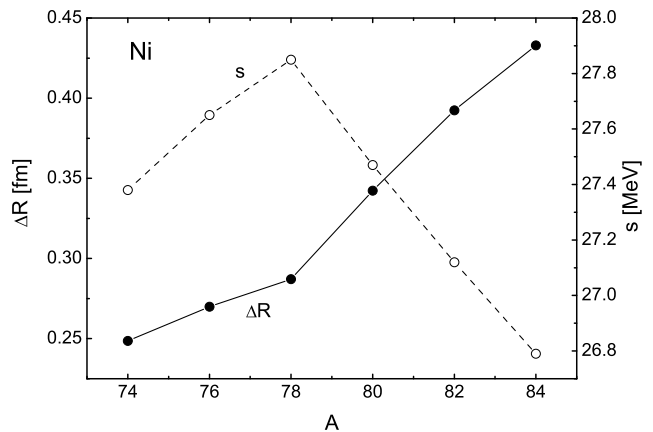}
\caption[]{HF+BCS neutron skin thicknesses $\Delta R$ (solid line)
and symmetry energies $s$ (dashed line) for Ni isotopes calculated
with SLy4 force.
\label{fig2}}
\end{figure}

\begin{figure}
\centering
\includegraphics[width=85mm]{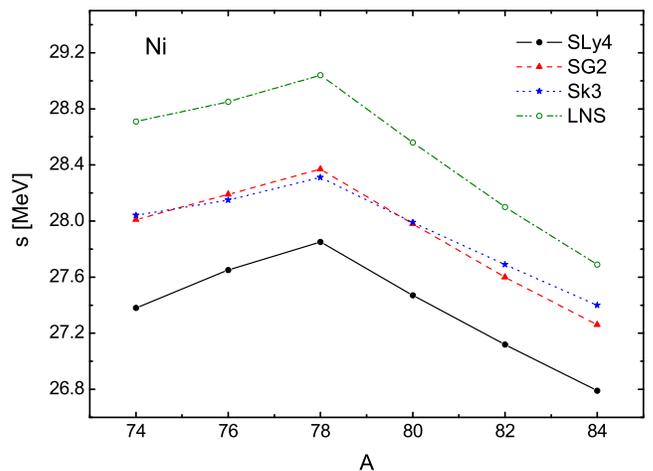}
\caption[]{(Color online) The symmetry energies $s$ for Ni isotopes calculated
with SLy4, SG2, Sk3, and LNS forces.
\label{fig3}}
\end{figure}

The results presented in Fig.~\ref{fig1}(a) are illustrated in
more details in Fig.~\ref{fig2}, where the evolution of both
neutron skin thickness and symmetry energy for Ni isotopes
calculated with SLy4 force is given simultaneously on the same
figure. For this chain our predictions for $\Delta R$ are extended
to more neutron-rich isotopes ($A=84$) comparing to the range of
isotopes considered in Ref.~\cite{Sarriguren2007}. The results for
the symmetry energy obtained from the four different Skyrme
parametrizations are compared in Fig.~\ref{fig3}. It is seen that
the trend of $s$ when the mass number increases is preserved for
all Skyrme forces but its values vary being very similar for SG2
and Sk3 forces. However, the magnitude of the symmetry energy
values slowly changes and turns out to be approximately in the
range of 27-29 MeV. The nuclear matter calculations relying on
realistic nucleon-nucleon interactions appear to yield volume
symmetry energy values within the same range of 27-29 MeV
depending on the interaction \cite{Engvik97}.

\begin{figure}
\centering
\includegraphics[width=80mm]{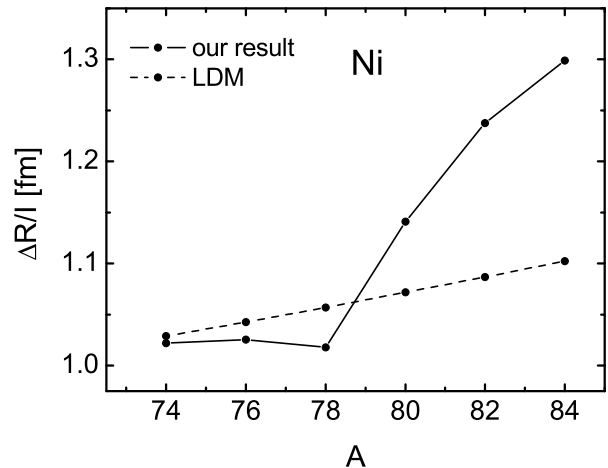}
\caption[]{The ratio $\Delta R/I$ for Ni isotopes calculated with
SLy4 force (solid line) and obtained from Eq.~(\ref{eq:35}) with
LDM \cite{Dan2003,Diep2007} (dashed line).
\label{fig4}}
\end{figure}

As is known, radioactive ion beams enable us to study radii of
unstable nuclei over a wide range of the relative neutron excess
(or relative neutron-proton asymmetry) $I=(N-Z)/A$. The role of
the global asymmetry on the neutron skin thickness in the Ni
isotopic chain is shown in Fig.~\ref{fig4} for the case of SLy4
force by presenting results for the ratio $\Delta R/I$ as a
function of the mass number $A$. It can be seen from
Fig.~\ref{fig4} that due to the gradual increase of the skin
thickness almost no dependence on $I$ is observed for isotopes
with $A<78$, while for the heavier ones the ratio $\Delta R/I$
starts to increase rapidly with $A$. The latter takes place namely
in the region where a pronounced neutron skin in Ni isotopes can
be expected. We also give in Fig.~\ref{fig4} estimations for this
ratio using the extended LDM formula for the binding energy of the
nucleus \cite{Dan2003,Diep2007} (with neglected Coulomb
contribution):
\begin{equation}
\Delta R^{LDM}=\frac{A^{2}R}{6NZ(1+A^{1/3}/y_{s})} I .
\label{eq:35}
\end{equation}
In Eq.~(\ref{eq:35}) $y_{s}\equiv S_{V}/S_{S}$ is the ratio
between the volume and surface symmetry energies and its value is
taken to be $y_{s}=1.7$ \cite{Diep2007}. For $R$ we use $R\approx
r_{0}A^{1/3}$ with $r_{0}=1.2$ fm.  As pointed out by Danielewicz
\cite{Dan2003}, Eq.~(\ref{eq:35}) is valid for the difference of
sharp-sphere radii, but it is generalized for the case of rms
radii that must be used. This creates an additional Coulomb
correction \cite{Dan2003} preserving unchanged the relation
between $\Delta R$ and $I$. We see that the difference between the
neutron and proton radii is linear in the global asymmetry in the
absence of the Coulomb contribution (which, in principle, cannot
be neglected) and measures the symmetry coefficient ratio. Thus,
Fig.~\ref{fig4} displays how much the mean-field results differ
from the LDM ones following the general observation that the
neutron skin depends roughly linearly on the asymmetry of the
nucleus. Although the Coulomb effects result in a reduction of the
neutron skin for $N>Z$ nuclei \cite{Dan2003,Diep2007}, it is
important to note the direct relation between the skin thickness
$\Delta R$ and the symmetry energy parameters $(S_{V},S_{S}$)
given by Eq.~(\ref{eq:35}).

In this aspect, here we would like to mention that the finite
nucleus symmetry energy within the extended LDM (which accounts
for the surface effects) is given by
\cite{Dan2003,Danielewicz,Diep2007}
\begin{equation}
a_{a}(A)=\frac{S_{V}}{1+y_{s}A^{-1/3}}.
\label{eq:35a}
\end{equation}
If one takes the same realistic values of $S_{V}=27$ MeV and
$y_{s}=1.7$ \cite{Diep2007} which we use when dealing with
Eq.~(\ref{eq:35}), it turns out that the values of $a_{a}$ are in
the interval 19--21 MeV for the nuclei ranging between $^{78}$Ni
and $^{208}$Pb. These values are significantly smaller than the
symmetry energy values $s$ derived in CDFM from Eq.~(\ref{eq:25}).
As far as one of the main goals of our work is to calculate the
nuclear symmetry energy of finite nuclei and to look for its
relationship with the neutron skin thickness in these nuclei from
a given isotopic chain, we would like to emphasize that the
results obtained by using Eq.~(\ref{eq:25}) represent estimations
of the symmetry energy for finite nuclei obtained on the basis of
nuclear matter EOS. A way to estimate the symmetry energy in
finite nuclei within the mean-field framework has been recently
proposed in \cite{Sama2007}, where Samaddar {\it et al.} studied
the density and excitation energy dependence of symmetry energy
and symmetry free energy of hot nuclei by using the local density
approximation.

In addition, it is important to note that by means of
Eq.~(\ref{eq:25}) we obtain the symmetry energy in finite nuclei
including both bulk and surface contributions on the base of the
Brueckner EOS. This is related to the usage in the method of the
weight function $|{\cal F}(x)|^{2}$ which is computed by taking
the derivative of the density distribution of a given finite
nucleus [Eq.~(\ref{eq:17})]. Thus, information from both parts
(central and surface) of the nuclear density is included in
$|{\cal F}(x)|^{2}$ and, consequently, in the total symmetry
energy. In this way, starting from the nuclear matter symmetry
energy, we calculate by our method the total symmetry energy for
the given finite nucleus. The use of different types of Skyrme
effective forces leads to density distributions of protons and
neutrons which slightly modify the weight function $|{\cal
F}(x)|^{2}$. Increasing the number of neutrons at fixed proton
number, Eq.~(\ref{eq:25}) provides values for the symmetry
energies $s$ of considered isotopes that are found to change
smoothly within a given chain.

As has been recently discussed (see, for instance,
Refs.~\cite{Dan2003,Diep2007, Nikolov2011}), the relative
contributions of the volume symmetry and surface symmetry energies
can be disentangled using the extended LDM formula for the binding
energy of the nucleus. A comparison between the symmetry energy
values obtained in \cite{Nikolov2011} (and listed in Table I of
this reference) for SLy4 parametrization and those calculated in
the present paper with the same force can be made. For example,
the values of the total symmetry energy for the case of SLy4
extracted from leptodermous expansion in Ref.~\cite{Reinhard2006}
and Eq.~(\ref{eq:35a}) vary between 19 and 23 MeV. Our values of
the total symmetry energy (deduced with $|{\cal F}(x)|^{2}$ from
calculations of the density corresponding to SLy4 force
[Eq.~(\ref{eq:17})]) lie in the range of 26.8--27.8 MeV for Ni,
27.3--28.8 MeV for Sn, and 28.7--29.1 MeV for Pb isotopes. In our
opinion, the differences of the compared results are due to the
peculiarities of the two methods. As noted in the previous
paragraph, in Eq.~(\ref{eq:25}) we use $|{\cal F}(x)|^{2}$
obtained from both parts (central and surface) of the nuclear
density, weighting by it the total symmetry energy for nuclear
matter $S^{NM}(x)$. Second, in the present work $S^{NM}(x)$ is
taken from the Brueckner theory, while in \cite{Nikolov2011}
(using the results of \cite{Reinhard2006}) the {\it bulk} symmetry
energy in the considered case of SLy4 parametrization is used. In
principle, other functionals apart from the Brueckner one reported
here can be applied within our method to calculate {\it the total
symmetry energy for finite nuclei}, but as described above, the
method gives it {\it unseparated into bulk and surface
contributions}.

\begin{figure*}
\centering
\includegraphics[width=160mm]{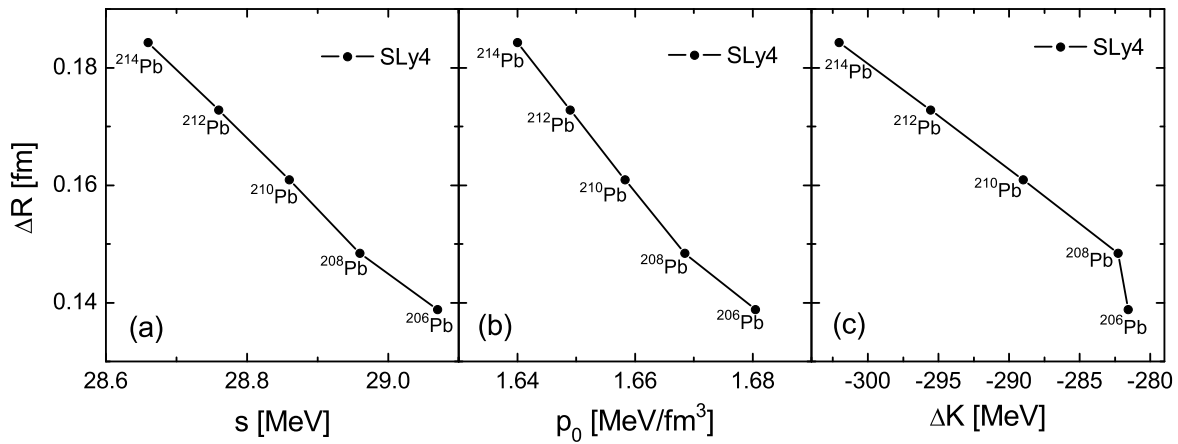}
\caption[]{HF+BCS neutron skin thicknesses $\Delta R$ for Pb
isotopes as a function of the symmetry energy $s$ (a), pressure
$p_{0}$ (b), and asymmetric compressibility $\Delta K$ (c)
calculated with SLy4 force.
\label{fig5}}
\end{figure*}

In Fig.~\ref{fig5} we display the theoretical neutron skin
thickness $\Delta R$ of nuclei from a Pb isotopic chain against
the parameters of interest $s$, $p_{0}$ and $\Delta K$. Similarly
to the three panels for Ni isotopes presented in Fig.~\ref{fig1},
the predicted correlations manifest the same linear dependence.
However, in this case the kink observed at double magic $^{208}$Pb
nucleus ($I \approx 0.21$) is much less pronounced that it can be
seen at $^{78}$Ni isotope ($I \approx 0.28$) and it does not
change its direction. The value of $\Delta R$ for $^{208}$Pb
deduced from present HF+BCS calculations with SLy4 force is lower
than the predicted thickness by the Skyrme Hartree-Fock model
\cite{Chen2005} and by the extended relativistic mean-field model
\cite{Agrawal2010}, but fits well the values calculated with
self-consistent densities of several nuclear mean-field models
(see Table I in Ref.~\cite{Centelles2010}). The $p_{0}$ and
$\Delta K$ values for $^{208}$Pb are in a good agreement with
those from Ref.~\cite{Chen2005}.

\begin{figure*}
\centering
\includegraphics[width=160mm]{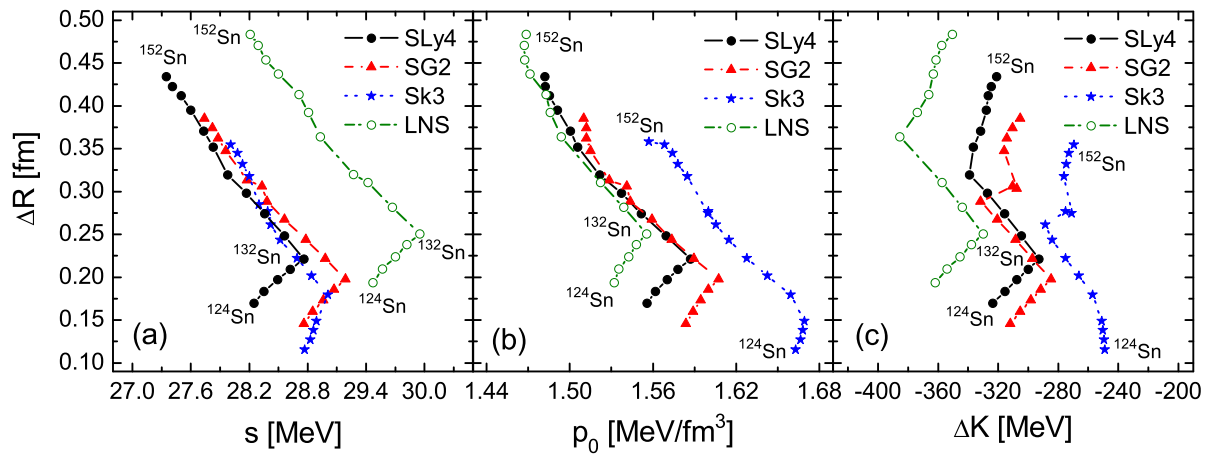}
\caption[]{(Color online) Same as in Fig.~\ref{fig1}, but for Sn isotopes.
\label{fig6}}
\end{figure*}

The analysis of the correlation between the neutron skin thickness
and some macroscopic nuclear matter properties in finite nuclei is
continued by showing the results for a chain of Sn isotopes. This
is done in Fig.~\ref{fig6}, where the results obtained with SLy4,
SG2, Sk3, and LNS Skyrme forces are presented for isotopes with
$A=124-152$. Similarly to the case of Ni isotopes with transition
at specific shell closure, we observe a smooth growth of the
symmetry energy till the double-magic nucleus $^{132}$Sn ($N=82$)
and then an almost linear decrease of $s$ while the neutron skin
thickness of the isotopes increases. In Ref.~\cite{Sarriguren2007}
we have studied a formation of a neutron skin in tin isotopes with
smaller $A$ where very poor experimental information is available.
For instance, a large uncertainty is shown to exist experimentally
in the neutron skin thickness of $^{124}$Sn, i.e., its value
varies from 0.1 to 0.3 fm depending on the experimental method.
Our theoretical prediction $\Delta R=0.17$ fm for this nucleus is
found to be within the above experimental band. A similar
approximate linear correlation between $\Delta R$ and $p_{0}$ for
Sn isotopes is also shown in Fig.~\ref{fig6}(b). Therefore, the
neutron skin thickness of these nuclei can be extracted once the
pressure $p_{0}$ at saturation density is known. The asymmetric
compressibility $\Delta K$ given in Fig.~\ref{fig6}(c) is less
correlated than $p_{0}$ with $\Delta R$ within the Sn isotopic
chains.

\begin{figure}
\centering
\includegraphics[width=85mm]{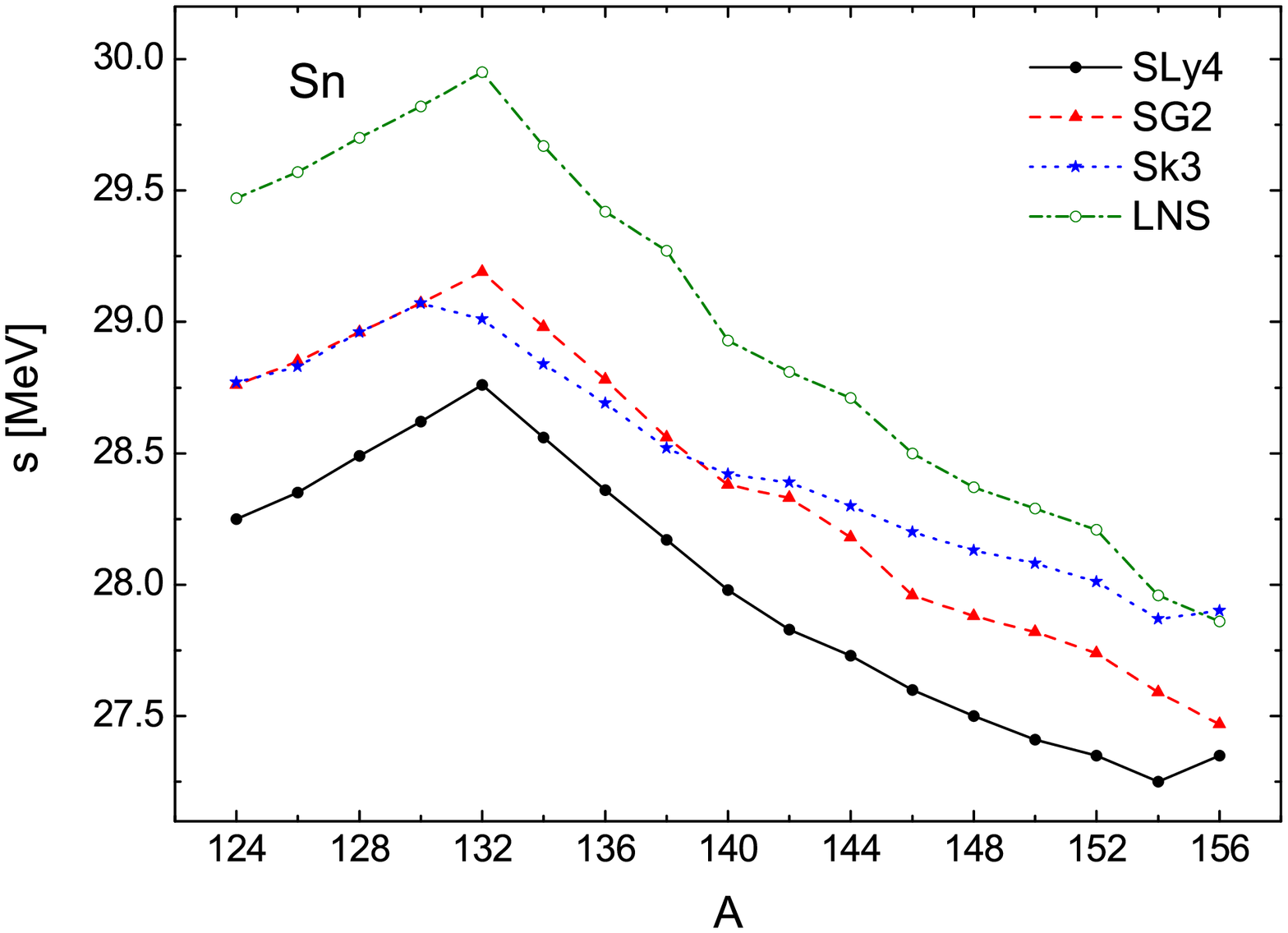}
\caption[]{(Color online) Same as in Fig.~\ref{fig3}, but for Sn isotopes.
\label{fig7}}
\end{figure}

\begin{figure}
\centering
\includegraphics[width=85mm]{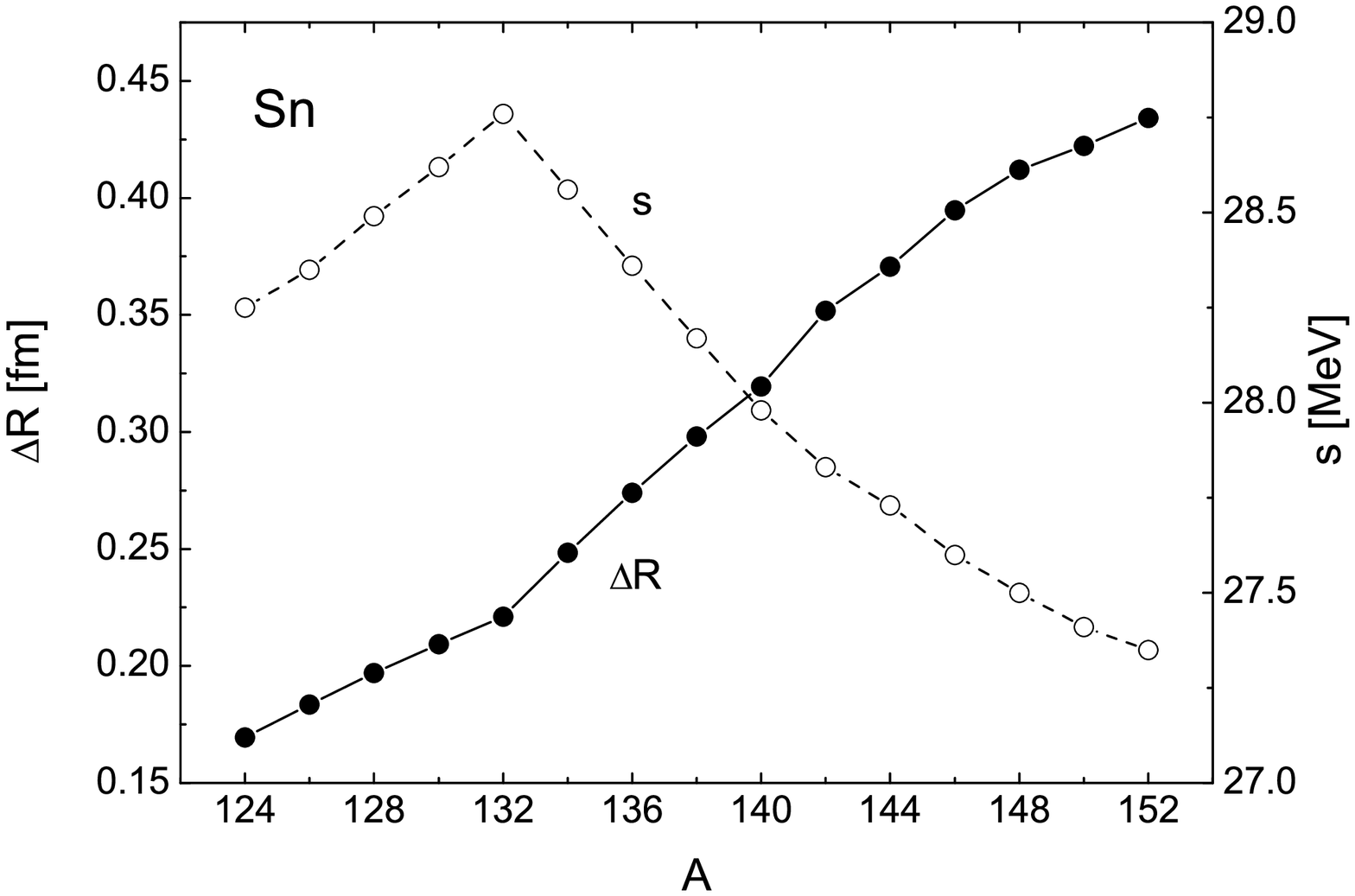}
\caption[]{Same as in Fig.~\ref{fig2}, but for Sn isotopes.
\label{fig8}}
\end{figure}

Figs.~\ref{fig7} and \ref{fig8} illustrate the evolution of the
symmetry energy with different Skyrme forces and the change of the
neutron skin for Sn isotopes in SLy4 case. It is seen from
Fig.~\ref{fig7} that the symmetry energies deduced from different
Skyrme parametrizations vary in the interval 27-30 MeV. Similarly
to the Ni isotopic chain (see Fig.~\ref{fig3}), the LNS force
yields values of $s$ that overestimate the corresponding values of
the other three Skyrme-type interactions. Also, the results for
the symmetry energies calculated with both SG2 and Sk3 forces are
comparable. At the same time the slope of all curves remains
almost constant. As can be seen from Fig~\ref{fig8}, the values of
$\Delta R$ for $^{130}$Sn and $^{132}$Sn obtained from our HF+BCS
calculations with SLy4 force fit very well the corresponding
neutron skin thicknesses of $0.23 \pm 0.04$ fm and of $0.24 \pm
0.04$ fm of these nuclei derived from pygmy dipole resonances
\cite{Klimk2007}. Moreover, both experimental and theoretical
values follow a trend established by a measurement of the stable
Sn isotopes \cite{Klimk2007,Sarriguren2007}. On the other hand,
the extended region of Sn isotopes in the present study gives a
good base for further measurements and systematic microscopic
model calculations in order to assess the isospin dependence of
the nuclear EOS.

\begin{figure}
\centering
\includegraphics[width=80mm]{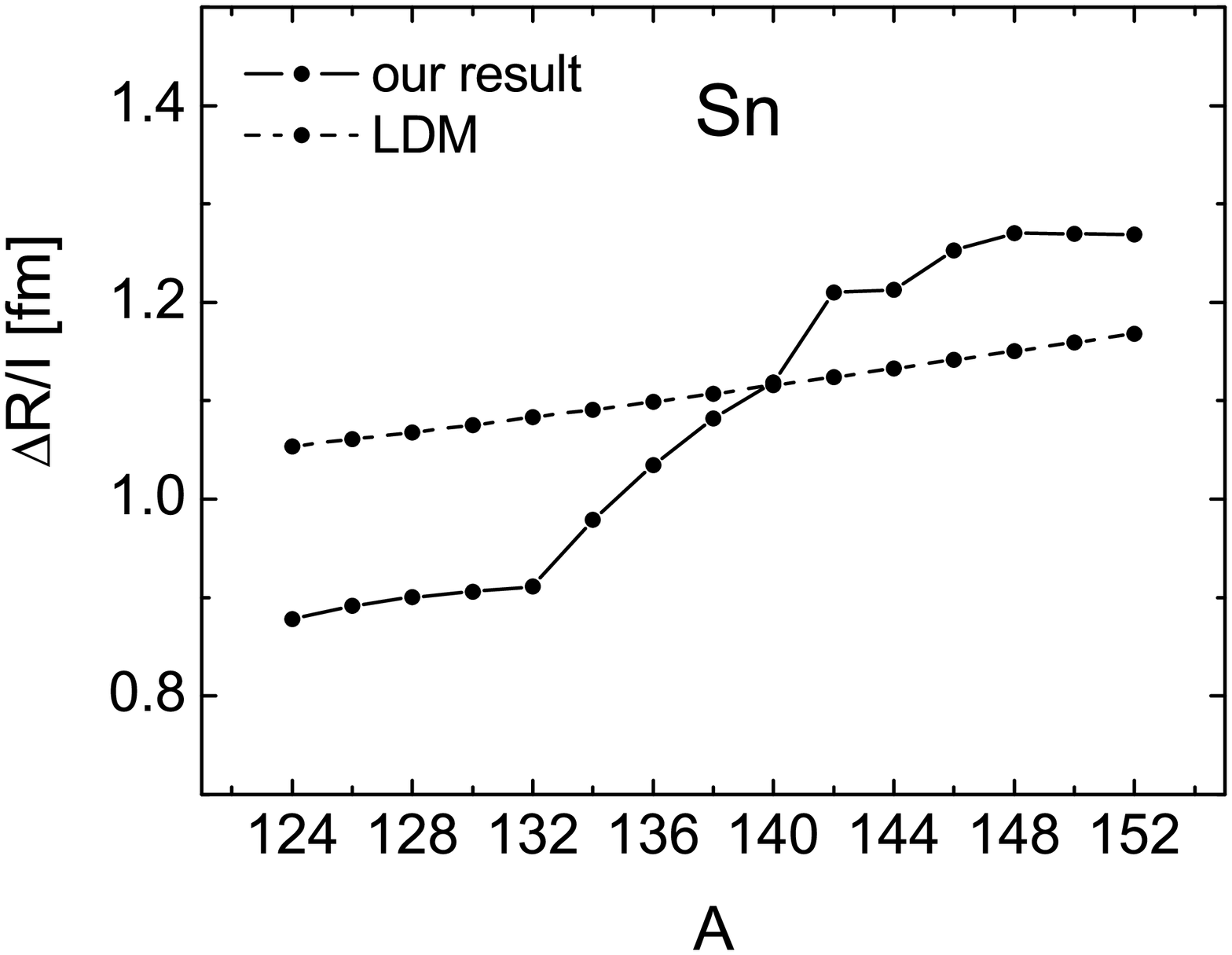}
\caption[]{Same as in Fig.~\ref{fig4}, but for Sn isotopes.
\label{fig9}}
\end{figure}

In Fig.~\ref{fig9} is given the evolution of the ratio $\Delta
R/I$ for Sn isotopes. A clear signal that a formation of a neutron
skin can be expected to start at $A>132$ is seen in favour of the
magic nature of $^{132}$Sn. Additionally, there is an indication
for a weak shell closure at $^{140}$Sn ($N=88$). As in the case of
Ni isotopes, the proportionality of the neutron skin thickness
$\Delta R$ to the asymmetry parameter $I$ following from the
liquid-drop mass formula [Eq.~(\ref{eq:35})] leads to a gradual
linear increase of $\Delta R/I$ with $A$ also within the Sn
isotopic chain. The corresponding numerical values of this ratio
turn out to occupy some intermediate range comparing to our HF+BCS
results.

The kinks displayed in Figs.~\ref{fig1}-\ref{fig3} by the Ni isotopes and in
Figs.~\ref{fig6}-\ref{fig8} by the Sn isotopes can be attributed to the shell structure
of these exotic nuclei. Indeed, the isotopic chains of Ni and Sn are of particular
interest for nuclear structure calculations because of their proton shell
closures at $Z=28$ and $Z=50$, respectively. They also extend from the proton drip
line that is found nearby the double-magic $^{48}$Ni (see, for instance, the
discussion in Ref.~\cite{Sarriguren2007} for a possible proton skin formation) and
$^{100}$Sn nuclei to the already $\beta$ unstable neutron-rich double-magic
$^{78}$Ni and $^{132}$Sn isotopes. In the case of $^{78}$Ni ($N=50$ shell closure)
the filling of the $1g_{9/2}$ orbit is completed. Beyond this isotope the
$2d_{5/2}$ subshell is being filled and a thick neutron skin is built up. Similar
picture is present for the Sn isotopes. Here, one finds a sudden jump beyond
$N=82$ where the $1h_{11/2}$ shell is filled and the $2f_{7/2}$ subshell becomes
populated. The situation in Pb isotopes shown in Fig.~\ref{fig5} is different from
those of Ni and Sn isotopes-no kinks appear in the Pb chain considered. In the case
of $^{208}$Pb ($N=126$) the $3p_{1/2}$ orbit is filled and above this nucleus
the first occupied level is $2g_{9/2}$. As a result, the bulk and surface contributions
to the neutron skin thickness of neutron-rich Sn and Pb isotopes reveal an
opposite effect, namely the surface part dominates the bulk one in tins, while for
Pb isotopes the bulk part is larger \cite{Warda2010}.

\begin{figure*}
\centering
\includegraphics[width=155mm]{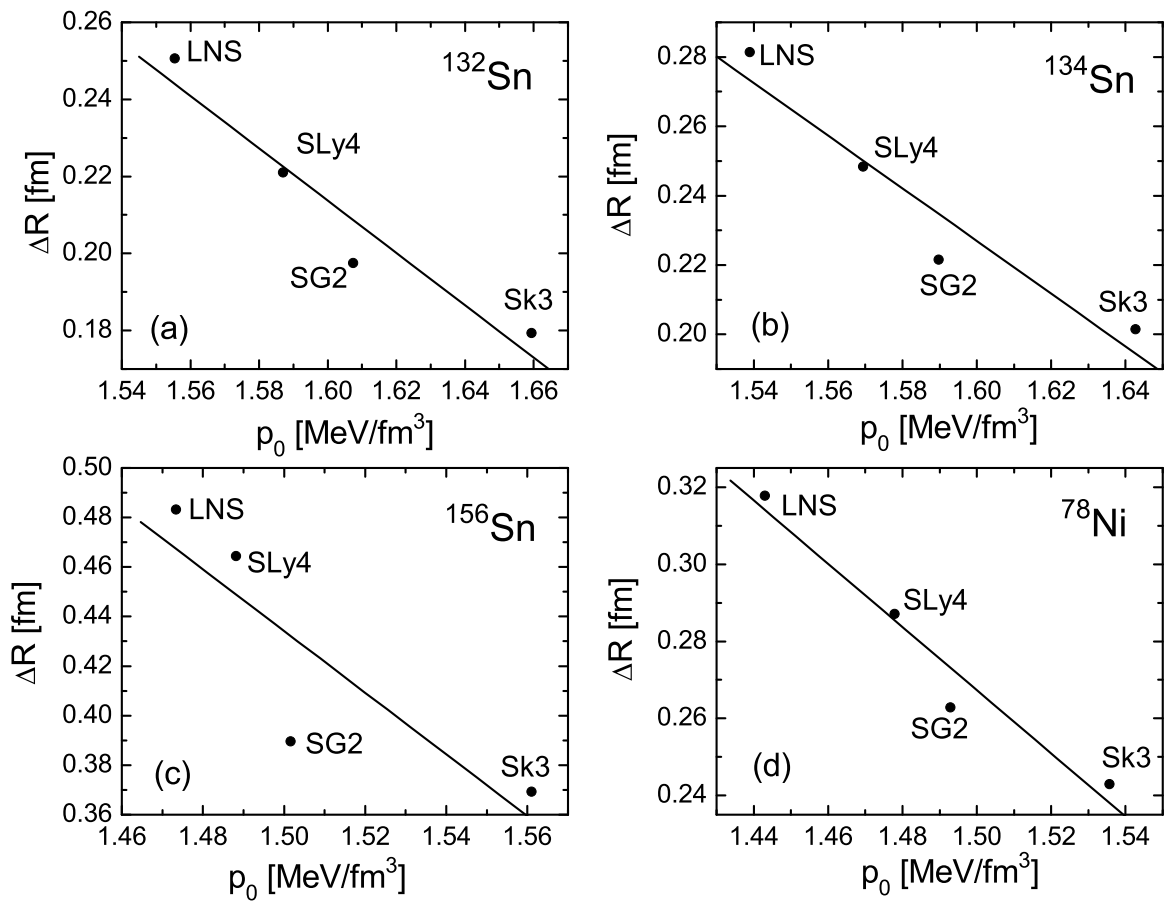}
\caption[]{HF+BCS neutron skin thicknesses $\Delta R$ for
$^{132}$Sn (a), $^{134}$Sn (b), $^{156}$Sn (c), and $^{78}$Ni (d)
as a function of the pressure $p_{0}$ for four sets of Skyrme
interaction parameters: SLy4, SG2, Sk3, and LNS. The lines in all
panels represent linear fits.
\label{fig10}}
\end{figure*}

As it has been shown by Brown and Typel (see
\cite{Brown2000,Typel2001}), and confirmed later by others
\cite{Steiner2005,Centelles2009,Chen2005,Furnstahl2002,Diep2003},
the neutron skin thickness (\ref{eq:34}) calculated in mean-field
models with either nonrelativistic or relativistic effective
interactions is very sensitive to the density dependence of the
nuclear symmetry energy and, in particular, to the slope parameter
L (respectively to the pressure $p_{0}$ [Eq.~(\ref{eq:6})]) at the
normal nuclear saturation density. We note that while the basic
idea behind studies of this type of correlations, namely between
the neutron skin thickness and the symmetry energy (or the slope
of the neutron EOS), is to constrain the symmetry energy in bulk
matter from experimental results in finite nuclei, the main aim of
our work is to extract quantitative information about $s$, $p_{0}$
and $\Delta K$ values for finite nuclei in CDFM. Moreover, so far
in our work we presented another type of correlation, namely how
these quantities are related to $\Delta R$ within a {\it given
isotopic chain} apart from the same one for a {\it specific
nucleus} using different theoretical models. Along this line we
analyze the latter correlation and results for several Sn
($^{132}$Sn, $^{134}$Sn, $^{156}$Sn) isotopes and $^{78}$Ni
nucleus are shown in Fig.~\ref{fig10}. Using the four sets of
Skyrme interaction parameters, a linear fit to the correlation
between $\Delta R$ and $p_{0}$ is performed to illustrate
qualitatively the above correlation. It is seen from
Figs.~\ref{fig10}(a) and \ref{fig10}(d) that in the cases of
double-magic $^{132}$Sn and $^{78}$Ni nuclei $\Delta R$ almost
linearly correlates with $p_{0}$. If one goes from $^{132}$Sn
nucleus further to unstable nuclei within the same Sn isotopic
chain the correlation becomes weaker, as it can be seen from
Figs.~\ref{fig10}(b) and \ref{fig10}(c), being almost the same in
the case of neighbouring $^{134}$Sn nucleus and poorly expressed
in the case of far away $^{156}$Sn isotope. Due to the limited
number of Skyrme parametrizations used in our theoretical
approach, the results presented in Fig.~\ref{fig10} cannot provide
a definite constraint of the slope ($p_{0}$ value) of the symmetry
energy. Obviously, more complete confirmation of the existing
correlation between the neutron skin thickness and the pressure
for a given nucleus can be drawn when large set of nuclear models
may be considered.

Finally, we would like to give estimations for the symmetry
energy, pressure and incompressibility  of asymmetric nuclear
matter corresponding to the Brueckner EOS provided by the
energy-density functional reported in
Refs.~\cite{Brueckner68,Brueckner69}. They are calculated at
equilibrium density $\rho_{0}=0.204$ fm$^{-3}$  for which the
energy per nucleon $V_{0}$ [Eq.~(\ref{eq:20})] has a minimum. The
latter ($E=-16.57$ MeV) is reached from the Brueckner formula at
Fermi momentum $k_{F}=1.446$ fm$^{-1}$ and at the value of
$\rho_{0}=0.204$ fm$^{-3}$ used by us, as shown in
Refs.~\cite{Bethe71,Brueckner68}. This value of $\rho_{0}$ is
considered to be quite realistic \cite{Bethe71}. The use of
Eqs.~(\ref{eq:24})--(\ref{eq:24b}) leads to the following
estimations: $S^{NM}=35.07$ MeV, $p_{0}^{NM}=1.82$ MeV/fm$^{3}$,
and $\Delta K^{NM}=-393.85$ MeV. It is seen that the symmetry
energy is few MeV above the experimental range 28 MeV $\leq S^{NM}
\leq 32$ MeV \cite{Moller95} and the values deduced from our
calculations for all finite nuclei considered. Also, the values of
$p_{0}^{NM}$ and $\Delta K^{NM}$ slightly overestimate the
corresponding ones obtained in finite nuclei. Actually, the
problem that the surface symmetry energy is very poorly
constrained in the current energy density functional
parametrizations, as well as by available phenomenological
estimates, has been recently indicated in Ref.~\cite{Nikolov2011}.
To illustrate this fact we would like to present here new symmetry
energy values derived from the extended LDM. By taking
$S_{V}=35.07$ MeV which, however, comes out from the Brueckner EOS
and a smaller value of $y_{s}=1.1$ (corresponding to the fit with
LDM formula in \cite{Diep2007}), Eq.~(\ref{eq:35a}) gives values
for the symmetry energy of 27.89 MeV for $^{78}$Ni, 28.84 MeV for
$^{132}$Sn and 29.58 MeV for $^{208}$Pb which are in a better
agreement with the results obtained from Eq.~(\ref{eq:25}).
Studying relationships between the neutron skin and various
properties of finite nuclei and infinite nuclear matter, it
follows from the comparison that our microscopic theoretical
approach produces results for these properties consistent with
those provided by the Brueckner energy-density functional which
may reduce the uncertainties of their future experimental
extraction.

\section{Conclusions}

In this work, a theoretical approach to the nuclear many-body
problem combining the deformed HF+BCS method with Skyrme-type
density-dependent effective interactions \cite{vautherin} and the
coherent density fluctuation model \cite{Ant80,AHP} has been used
to study nuclear properties of finite nuclei. For this purpose, we
examined three chains of neutron-rich Ni, Sn, and Pb isotopes,
most of them being far from the stability line and representing an
interest for future measurements with radioactive exotic beams.
Four Skyrme parametrizations were involved in the calculations:
SG2, Sk3, SLy4, and LNS. In addition to the interactions used in
Ref.~\cite{Sarriguren2007}, the LNS effective interaction was used
in the present paper, because it has been built up from the EOS of
nuclear matter including effects of three-body forces
\cite{Cao2006}.

For the first time, we have demonstrated the capability of CDFM to
be applied as an alternative way to make a transition {\it from
the properties of nuclear matter to the properties of finite
nuclei} investigating the nuclear symmetry energy $s$, the neutron
pressure $p_{0}$ and the asymmetric compressibility $\Delta K$ in
finite nuclei. This has been carried out on the base of the
Brueckner energy-density functional for infinite nuclear matter.
Mainly, the symmetry energy has been determined from analyzing
nuclear masses within liquid-drop models, while much more effort
has been recently devoted to extracting the value of its slope
parameter. Instead of this, in the present work we applied the
CDFM scheme. One of the advantages of the CDFM is the possibility
to obtain transparent relations for the intrinsic EOS quantities
analytically by means of a convenient approach to the weight
function. The key element of this model is the choice of density
distributions which were taken from self-consistent deformed
HF+BCS calculations.

We would like to emphasize that our new method allows one to
estimate the symmetry energy and, therefore, the symmetry-energy
coefficient in finite nuclei thus avoiding the problems related to
fitting the Hartree-Fock energies to a LDM parametrization.

We have found that there exists an approximate linear correlation
between the neutron skin thickness of even-even nuclei from the Ni
($A=74-84$), Sn ($A=124-152$), and Pb ($A=206-214$) isotopic
chains and their nuclear symmetry energies. Our HF+BCS
calculations lead to a symmetry energy in the range of 27-30 MeV,
which is in agreement with the empirical value of $30\pm 4$ MeV
\cite{Haustein88} and with the results of Nikolov {\it et al.}
\cite{Nikolov2011}. In the cases of Ni and Sn isotopes the
symmetry energy is found to increase almost linearly with the mass
number $A$ till the double-magic nuclei ($^{78}$Ni and $^{132}$Sn)
and then to decrease, being larger when LNS force is used. A
similar linear correlation between $\Delta R$ and $p_{0}$ is also
found to exist, while the relation between $\Delta R$ and $\Delta
K$ is less pronounced. The same behavior containing an inflexion
point transition at specific shell closure is observed for these
correlations. The calculated values of $p_{0}=1.36-1.68$
MeV/fm$^{3}$ obtained in our calculations lead to values of the
slope parameter $L=26-32$ MeV that are in agreement with other
theoretical predictions (see, for example \cite{Chen2005}). Hence,
the study of other two EOS parameters (whose determination is more
uncertain than the symmetry energy), namely the slope and the
curvature of $s$, performed in our work may provide more
constraints on both the density dependence of the nuclear symmetry
energy and the thickness of the neutron skin in neutron-rich
medium and heavy nuclei.

In the present work, the role of the relative neutron-proton
asymmetry $I$ on the size of the neutron skin $\Delta R$ was
studied for Ni and Sn isotopes. For this purpose the results for
the ratio $\Delta R/I$ were presented and a comparison with its
estimation using the extended liquid-drop mass formula
[Eq.~(\ref{eq:35})] was made. The HF+BCS results show that a
formation of a neutron skin can be expected to start at $A>78$ and
$A>132$ for Ni and Sn isotopes, respectively. Although the
proportionality that exists between $\Delta R$ and $I$ in
(\ref{eq:35}) produces smooth change of $\Delta R$ with $A$, the
size of the neutron skin in both cases is consistent with the
results of other calculations with the SLy4 parametrization
\cite{Mizutori2000}, as well as with results of spherical
self-consistent HFB calculations with finite-range forces of the
Gogny type \cite{Schunck2008}. It is confirmed that the neutron
skins in nuclei with a large neutron excess near and beyond the
drip line are very clearly related to the asymmetry parameter $I$.
However, the shell effects, which are always present in mean-field
calculations, produce deviations from the linear dependence
between $\Delta R$ and $I$ in these nuclei.

Summarizing, in this work we use a microscopic theoretical
approach to study important macroscopic nuclear matter quantities
in finite nuclei and their relation to surface properties of
neutron-rich exotic nuclei. We have shown that this approach gives
a very reasonable description of the neutron skin thicknesses of
several Sn isotopes which have been experimentally measured. A
good agreement is achieved with other theoretical predictions for
the volume symmetry energies of Ni isotopes, as well as for the
pressure and asymmetric compressibility of nuclei within the Pb
isotopic chain. The capability of the present method can be
further demonstrated by taking into consideration Skyrme-type and
relativistic nuclear energy-density functionals. In this way, more
systematic analysis of the relationship between the symmetry
energy and the neutron skin thickness will be carried out and more
definite conclusions on its isotopic sensitivity will be drawn.

\begin{acknowledgments}
Two of the authors (M.K.G. and A.N.A.) are grateful for the
support of the Bulgarian Science Fund under Contract No.~02--285.
E.M.G. and P.S. acknowledge support from MICINN (Spain) under
Contracts FIS2008--01301 and FPA2010--17142.
\end{acknowledgments}

\end{document}